%% file: milliqan.tex
\documentclass[
superscriptaddress,
 amsmath,amssymb,
 aps,
prl,twocolumn,
floatfix,
]{revtex4-2}
\usepackage{graphicx}
\usepackage{dcolumn}
\usepackage{bm}
\usepackage{xcolor}
\usepackage{xspace}
\usepackage{topcapt}
\usepackage{lipsum}

\usepackage{multirow}

\newcommand{\mCP}{\ensuremath{\chi}\xspace}
\newcommand{\mCPp}{\ensuremath{\mCP^+}\xspace}
\newcommand{\mCPm}{\ensuremath{\mCP^-}\xspace}

\begin{document}

\preprint{CERN-EP-2025-120}

\title{Search for millicharged particles in proton-proton collisions at $\sqrt{s} = 13.6$~TeV}

\include{authorlist}

\date{\today}

\begin{abstract}
\noindent We report on a search for elementary particles with charges much smaller than the electron charge using a data sample of proton-proton collisions provided by the CERN Large Hadron Collider in 2023--24, corresponding to an integrated luminosity of 124.7~fb$^{-1}$ at a center-of-mass energy of 13.6~TeV. The analysis presented uses the completed Run 3 milliQan bar detector to set the most stringent constraints to date for particles with charges $\leq0.24~\rm{e}$ and masses $\geq0.45~\rm{GeV}$. 
\end{abstract}

\maketitle

There are strong experimental and theoretical motivations for the existence of a ``dark sector" containing one or more types of particles that could compose the non-luminous dark matter hypothesized to explain gravitational effects observed in astrophysics and cosmology~\cite{ArkaniHamed:2008qn, Pospelov:2008jd,Farzan:2019qdm,Foot:2007cq,Wallemacq:2014sta,Gies:2006ca,Gan:2023jbs}. Consequently, a number of experiments have been performed at various terrestrial facilities to search for direct evidence of such a dark sector~\cite{Battaglieri:2017aum, Beacham:2019nyx,Strategy:2019vxc}. One type of potential dark sector constituent, which may or may not be the galactic dark matter, that has grown in importance over the last decade is a particle with electric charge much smaller than that of the electron. Such particles have come to be called ``millicharged particles" (mCPs) in the literature, possibly because a natural value of $Q \sim \alpha \rm{e}/  \pi \approx 0.001\ \rm{e}$ arises from one-loop effects in models of this type~\cite{Davidson:1993sj}.

Owing to their feeble electromagnetic interaction, mCPs are essentially invisible to ionization-based detectors designed for the comparatively large energy deposition of a minimum ionizing particle (MIP). Consequently, direct detection of mCPs requires dedicated detectors or re-purposed detectors that were designed for other kinds of feebly interacting particles (e.g. neutrinos). While previous experiments have searched for mCPs for some time~\cite{Prinz:1998ua,Essig:2013lka,CMS:2012xi,Davidson:2000hf, Badertscher:2006fm,Magill:2018tbb}, in the decade since the publication of~\cite{Haas:2014dda} by two of the authors of this paper, which led to the formation of the milliQan collaboration~\cite{Ball:2016zrp}, the number of experiments that have either directly searched for, re-analyzed their data to search for, or are proposed to search for, mCPs has proliferated. 
At accelerators, those who have recently set direct constraints on the parameter space spanned by the mass and charge of the mCP in addition to milliQan~\cite{Ball:2020dnx} include CMS~\cite{CMS:2024eyx}, MilliQ~\cite{Prinz:1998ua}, MiniBooNE~\cite{Magill:2018tbb,MiniBooNEDM:2018cxm}, LSND~\cite{Magill:2018tbb,LSND:2001akn}, ArgoNeuT~\cite{Harnik:2019zee,ArgoNeuT:2019ckq}, and SENSEI~\cite{SENSEI:2023gie}. Indirect constraints on this parameter space have also been set by astrophysical, cosmological, and non-accelerator based terrestrial experiments~\cite{Davidson:1991si,Mohapatra:1990vq,Davidson:1993sj,Davidson:2000hf,CDMS:2014ane,Emken_2019,PhysRevD.99.032009,PhysRevLett.120.211804,PhysRevLett.99.161804,PhysRevLett.113.251801,Brust:2013ova,Vogel_2014,Plestid:2020kdm,Kachelriess:2021man,ArguellesDelgado:2021lek,Harnik:2020ugb,Fung:2023euv}. However, such constraints can be easily evaded by adding extra degrees of freedom to dark sector models~\cite{Izaguirre:2015eya}.

The milliQan experiment is sited in the PX56 drainage gallery at LHC Point 5. The nominal center of milliQan is at a radial distance of 33~m from the CMS Interaction Point (IP) with 17~m of rock between the CMS IP and the experiment, which provides shielding from most particles produced in LHC collisions. The milliQan Run 3 detector consists of two scintillator arrays aligned using standard laser-based survey techniques such that the center of the scintillator arrays project back to within 1~cm of the CMS IP. The two arrays employ different shapes of scintillator to provide complementary sensitivity to the mCP parameter space~\cite{milliQan:2021lne}. Echoing their respective geometries, we refer to them as the ``bar" and ``slab" detectors. In the CMS coordinate system~\cite{CMSTDR}, the bar detector is positioned at an azimuthal angle ($\phi$) of $43^\circ$ and pseudorapidity ($\eta$) of $0.1$. The slab detector is located 5~m behind the bar detector at $\phi = 38^\circ$. The two arrays were installed sequentially with the bar detector completed in mid 2023 and the slab detector completed in early 2025. In this Letter, we present the analysis of proton-proton collision data provided by the CERN Large Hadron Collider (LHC) in 2023--24 that was recorded by the Run 3 milliQan bar detector only. This data sample corresponds to an integrated luminosity of $124.7 \pm 3.8$~fb$^{-1}$ at a center-of-mass energy of 13.6~TeV.

The milliQan Run 3 bar detector consists of 64 EJ-200~\cite{Eljen} plastic scintillator bars arranged in four radial ``layers" with $4 \times 4$ bars per layer. Each bar is $60 \times 5 \times 5\ \rm{cm}^3$, with the long dimension pointing to the CMS IP. Additional scintillator components are deployed to provide discrimination against two background processes. Three thin scintillator panels surround the ``top" and ``sides" of the bar array in order to provide an active veto of particles from cosmic rays. Two thin scintillator panels are also put on either ``end" of the array in order to aid in the vetoing of through-going MIPs (e.g. muons produced in LHC collisions). 

Scintillation light produced in the bars/panels is detected by Hamamatsu R878 ~\cite{Hamamatsu} photomultiplier tubes (PMTs) optically coupled to the scintillator volumes. The PMT anode current is amplified by a custom circuit board, built into the PMT base, giving typical single-photoelectron (SPE) pulses with $\approx 20$ mV amplitude. The output waveforms are readout via five 16-channel CAEN V1743 digitizers~\cite{CAENV1743} with a 2.56~$\mu$s acquisition window. Trigger requests are generated when a pulse in either of two channels in a ``trigger group" exceeds a set threshold, chosen to be fully efficient for SPE pulses, that ranges from 12--18~mV for channels connected to scintillator bars. These LVDS signals are input into a custom FPGA trigger board that contains a menu of logic conditions implemented in firmware that allows up to 16 independent readout decisions to be made in real time.

Recorded data consists of waveforms of voltage against time for each channel. In order to reconstruct the passage of particles through the scintillator, ``pulses" are reconstructed from each waveform. The efficiency for each PMT to reconstruct scintillator photons produced by energy deposited in their corresponding bar is calibrated using 22~keV X-rays from a $^{109}\rm{Cd}$ radioactive source. A high purity sample of through-going muons originating from LHC collisions is used to calibrate the arrival time of pulses to account for differences in time due to signal propagation through the detector medium, cables of varying lengths, and electronics. Timing is corrected such that particles traveling through the detector have calibrated arrival times that all occur at the same time. We also use this sample of through-going muons originating from LHC collisions to check the alignment of the milliQan bar detector. By comparing the entry and exit positions of such muons, we find the bar detector to be well aligned horizontally, but with a slight vertical misalignment of $-0.2^{\circ}$ relative to nominal. This misalignment leads to a $\sim 12\%$ reduction in signal efficiency.

To optimize milliQan's sensitivity to mCPs over the largest range of the charge-mass plane we perform detailed simulations of the production mechanisms that could result in mCPs at the LHC, as well as the dominant background processes that mimic the mCP signal. We generate mCPs produced through the Drell-Yan process, as well as from $\Upsilon$, J/$\psi$, $\psi$(2S), $\phi$, $\rho$, and $\omega$ decays into $\mCPp\mCPm$, and from Dalitz decays of $\pi^0$, $\eta$, $\eta'$, and $\omega$ following the procedure described in our 13~TeV search~ \cite{Ball:2020dnx}, with the following differences. Events were regenerated in \textsc{Pythia8}~\cite{Sjostrand:2007gs} and \textsc{MadGraph5}\_a\textsc{mc@nlo}~\cite{Alwall:2014hca} with the $pp$ collision energy increased from 13 to 13.6~TeV. The parton distribution functions used to generate mCPs coming from charmonium produced in B meson decays was changed from \texttt{CTEQ6.6} to \texttt{NNPDF30\_NLO\_as0118} in \texttt{FONLL}~\cite{Cacciari:1998it}. For direct charmonium we used theoretical calculations at $\sqrt{s}=13.6$~TeV~\cite{Ma:2010jj,Ma:2010yw,Ma:2014mri}. For direct bottomonium, the same fits to the LHCb production cross section ratios that were used to scale from 7 to 13 TeV were used to scale from 13 to 13.6 TeV~\cite{Sirunyan:2017qdw,Aad:2011xv,Aad:2012dlq,Aaij:2018pfp}. In addition to these effects corrections were made to the normalizations of direct charmonium and bottomonium production to account for the fact that in~\cite{Ball:2020dnx} pseudorapidity was used in place of rapidity in a portion of phase space where the approximation was not valid.

Generated mCPs are propagated from the CMS IP to 2~m from the face of the bar detector taking into account magnetic deflection, multiple scattering, and energy loss, after which they are fed into a full \textsc{Geant4}~\cite{GEANT4:2002zbu} simulation that includes a geologically accurate model of the remaining rock, the geometry of the drainage gallery, and a detailed component level description of the Run 3 bar detector as summarized in Ref~\cite{milliQan:2021lne}. The mCP interactions are explicitly modeled and contain ionization energy deposits, scaled by $Q^2$, and incorporate multiple scattering. In order to realistically replicate the effects of noise and electronics on the detector response, we use templates of pulses extracted from data to inject waveforms into the simulation. To validate our simulation, we measure the rate of through-going muons originating from LHC collisions, and compare the obtained flux to the rate predicted from the simulation. From 124.7 fb$^{-1}$ $pp$ collision data, we identify 19955 particles that have the trajectory, pulse area, and timing consistent with muons originating from the CMS IP for a measured flux of $0.160 \pm 0.010~ \text{muons}/$pb$^{-1}$. The corresponding flux predicted for such muons in Monte Carlo simulation is $0.22\pm0.06~$\text{muons}/pb$^{-1}$, dominated by uncertainties in low momentum production modes~\cite{Khachatryan_2017}. 

\begin{table*}[ht]
    \centering
    \resizebox{7in}{!}{%
        \begin{tabular}{l|ccc|ccc}
    \hline 
     \hline 
     \multirow{4}{*}{Selection Criteria} & \multicolumn{3}{c|}{Signal Region 1} & \multicolumn{3}{c}{Signal Region 2} \\ 
     \cline{2-7} 
     & Data &  Signal & Signal & Data & Signal & Signal \\ 
     & Beam-On & m=0.1 GeV & m=1.0 GeV & Beam-On & m=1.7 GeV & m=10.0 GeV \\ 
     & t=3393~h & Q/e=0.004 & Q/e=0.008 & t=3393~h & Q/e=0.03 & Q/e=0.2 \\ 
     \hline 
    Triggered Events & 26864552 & 324.0  & 61.3 & 26864552 & 27.0 & 37.2\\ 
    Cosmic Muon Veto & 790776 & 324.0  & 61.3 & 790776 & 27.0 & 37.2\\ 
    Pulse/Event Quality & 506417 & 323.9  & 61.3 & 790383 & 27.0 & 37.2\\ 
    Shower Veto & 3369 & 12.0  & 19.3 & 9152 & 7.7 & 9.5\\ 
    {\bf SR1}: $\leq 4$ Bars & 985 & 11.7  & 19.3 & --- & --- & ---\\ 
    Noise Filter & 985 & 11.7  & 19.3 & 9113 & 7.7 & 9.5\\ 
    Energy Max/Min & 336 & 10.3  & 16.5 & 1827 & 7.6 & 9.5\\ 
    {\bf SR1}: Beam Muon Veto & 331 & 10.3  & 16.5 & --- & --- & ---\\ 
    {\bf SR1}: End Panel Veto & 209 & 10.1  & 14.3 & --- & --- & ---\\ 
    Straight Line & 3 & 9.2  & 14.3 & 1372 & 7.5 & 9.4\\ 
    $\Delta$ T(max-min) $\leq 20$ ns & 0 & 8.7  & 14.1 & 1355 & 7.5 & 8.6\\ 
    {\bf SR2}: End Panel Required & --- & ---  & --- & 1320 & 5.8 & 8.2\\ 
    {\bf SR2}: $\leq 4$ Bars & --- & ---  & --- & 84 & 5.8 & 7.3\\ 
    {\bf SR2}: $\text{nPE}_{\text{max}}^{\text{Panel}} < 70$ & --- & ---  & --- & 2 & 5.8 & 7.0\\ 
    \hline 
     \hline
    \end{tabular}
    }    
    \caption{Sequential impact of selection criteria on the number of events in the mCP search. Criteria in same row can differ between SR1 and SR2 as detailed in text. Bold type indicates criteria that are applied only to SR1 or SR2.}
    \label{tab:cutflow}
\end{table*}

The 2023--24 dataset analyzed for this search was collected over periods where the LHC was colliding protons (``beam-on'' data) and periods with no collisions (``beam-off'' data). We use beam-on data to search for the mCP signal, and beam-off data as a statistically independent sample of events containing background processes. Data were collected on two unprescaled signal triggers implemented in the trigger board FPGA: one requiring aligned hits across three layers that point to the CMS IP, and another requiring hits in four detector layers. Additional triggers were used to collect data from background sources. Total trigger live-times were 3393 and 4471 hours for the beam-on and beam-off data sets, respectively. The average total trigger rate over this time period was $\approx2.2$~Hz. We search for an mCP signal in this dataset by selecting events with the signature of a pulse in each of the four layers of the Run 3 milliQan bar detector in the channel corresponding to same row and column of the array in each layer. Non-mCP background processes can mimic this signature: PMT pulses that are not due to a photon impinging on the cathode such as dark current arising from thermionic emission of electrons, or afterpulses of ionized residual gas in the tube; neutral and charged particles in showers caused by muons from either cosmic rays or LHC collisions; and radiation in the cavern, scintillator bars, or surrounding material. Since the timing of pulses due to dark current is random, this background is suppressed by requiring a four-fold coincidence within a timing window of 20~ns between each of the four layers. This threshold is chosen to be $>90\%$ efficient for all masses and charges.

We search for mCPs in two distinct signal regions designed to target mCPs with different charges. The first signal region (SR1) targets charges from $10^{-3} \lesssim Q/e \lesssim 10^{-2}$ while the second signal region (SR2) targets charges from $10^{-2} \lesssim Q/e \lesssim 10^{-1}$. Candidate events in both signal regions are required to pass selection criteria that enforce data quality by ensuring digitizer synchronization and rejecting electronic noise. We reject afterpulses by selecting the first pulse in an event within the trigger window. We reject events coming from cosmic muons or their associated showers by requiring there are no hits in the top or side panels. We also place an upper limit on the ratio of the max/min energy deposited in the 4 bar channels. The max/min energy ratio is required to be $< 5$ if any pulse has an energy $>50$~keV. Otherwise the max/min ratio is set at $<10$. Finally, we select events with at least one hit per layer.

In addition to these common selections, we apply selection criteria specific to SR1 and SR2. For SR1, we select events that 
travel straight through our detector depositing small amounts of energy in all bars. Events are required to contain exactly four bar pulses, contain no pulses in the end panels (front and back), and no pulses with an energy deposition that saturates the readout electronics, which corresponds to $\sim350$ photoelectrons. Finally, we require that selected events have hits in the same row/column in all four layers, and that the time difference between the first and last hit is $<20$~ns, as previously discussed. The full set of criteria and their impact on the number of selected events in SR1 are given in Table~\ref{tab:cutflow}.

In SR2, we target larger charges for which we expect sufficient energy deposition in the thin panels to efficiently reconstruct pulses. We therefore require there is at least one hit in either end panel. Similar to SR1, we require that there are 4 hits straight through the detector, and that those 4 hits have a maximum timing difference of 20~ns. In SR2, however, the dominant background contribution are muons originating from LHC collisions. In order to effectively veto these muons we require that exactly 4 bars are hit and that any pulses in the end panels are $<70$ photoelectrons. These criteria are used because beam muons on average deposit more than 70 photoelectrons in the end panels and can cause hits in more than 4 bars. The full set of criteria and their impact on the number of selected events in SR2 are given in Table~\ref{tab:cutflow}.

We estimate the residual background contamination in SR1 using an ``ABCD" method. We divide the plane given by the timing difference between the first and last hit in an event ($\Delta T_{max,min}$) on the vertical axis and the requirement that pulses align in a straight line (i.e. that the channels containing the pulses are in the same position within each layer such that they point to the CMS IP) on the horizontal axis, into four quadrants. In this plane the mCP signal region is the lower-right quadrant that contains events that pass the straight-line requirement and have a $\Delta T_{max,min} < 20$~ns. We label this region as ``A". The remaining quadrants are labeled clockwise from this region as ``B", ``C", and ``D". Assuming these are independent variables, the background contribution to the signal region is extrapolated by the algebraic expression ``BD/C." We validate this assumption using the beam-off data control sample in which the background contribution to the signal region is predicted to be $0.32^{+0.24}_{-0.16}$. We observe 0 events in the signal region of the beam-off dataset, consistent with the ABCD background estimate for SR1. We further validate the assumption of independence using a “nearly-pointing” control sample in the beam-on dataset, consisting of events with straight-line trajectories that shift by one row or column. We observe 0 events in region A of this control sample, while the calculation BD/C predicts $0.31^{+0.28}_{-0.18}$. The uncertainty in the number of predicted background events in SR1 is dominated by the limited statistics in the regions used to make the estimate.

With our background estimation method validated, we use it to predict a background of $0.10^{+0.12}_{-0.07}$ events in SR1. Finally, we search the beam-on dataset for mCPs in SR1. Out of 26,864,552 triggered events collected while LHC proton beams were colliding, we observe 0 candidate mCP events passing all selection criteria for SR1.

Similarly, we estimate the residual background contamination in SR2 using another ``ABCD" method. In this case we divide the plane given by the maximum number of photoelectrons in the end panels on the horizontal axis and the number of bars hit on the vertical axis. These variables are chosen because in SR2 the dominant background source is muons from LHC collisions. Such muons deposit significant energy in the end panels, and shower in the detector resulting in a large number of bars hit. In this plane the mCP signal region is the lower-left quadrant that contains events that have few photoelectrons in the end panels and few bars hit. We label this region as ``A". The remaining quadrants are labeled counter-clockwise from this region as ``B", ``C", and ``D". As in the case for SR1, the background contribution to signal region for SR2 is extrapolated by the algebraic expression ``BD/C." 

From the beam-off dataset, we know the dominant background contribution to SR2 is muons from LHC collisions so it is not a useful control sample to validate the SR2 background estimation method. Instead, we check the validity of the SR2 background estimate with a closure test performed in the sample formed by events populating the background dominated regions C and D. From these two regions, a new, independent ABCD plane is constructed by selecting events with more than 4 bars hit. Within this subset, we redefine the ABCD regions: the new region A (treated as the signal region for this closure test) is defined by events with $\leq 6$ bars hit and a maximum panel nPE $\leq 70$. The estimated contribution to the new region A is $3.40^{+1.69}_{-1.20}$, which is consistent with the observed count of 5 events.

With our background estimation method validated, we use it to predict a background of $0.87_{-0.26}^{+0.33}$ in SR2. As for SR1, the uncertainty in the number of predicted background events in SR2 is also dominated by the limited statistics in the regions used to make the estimate. Out of 26,864,552 triggered events collected while LHC proton beams were colliding, we observe 2 candidate mCP events passing all selection criteria for SR2.

A summary of the results of the observed events in SR1 and SR2 compared to their respective background contribution predictions is given in Table~\ref{tab:bkgPredbeam}.


\setlength{\textfloatsep}{5pt}

\begin{table}[ht]
\renewcommand{\arraystretch}{1.2}
\centering
\begin{tabular}{cccc}
\hline
\hline
Signal Region & Prediction & Observation\\
\hline
SR1 & $0.10^{+0.12}_{-0.07}$ & 0\\
SR2 & $0.87_{-0.26}^{+0.33}$ & 2\\
\hline
\hline
\
\end{tabular}%

\caption{Summary of the results of the search for mCPs. \label{tab:bkgPredbeam}}
\end{table}

\begin{figure}[!h]
     \centering 
     \includegraphics[width=\columnwidth]{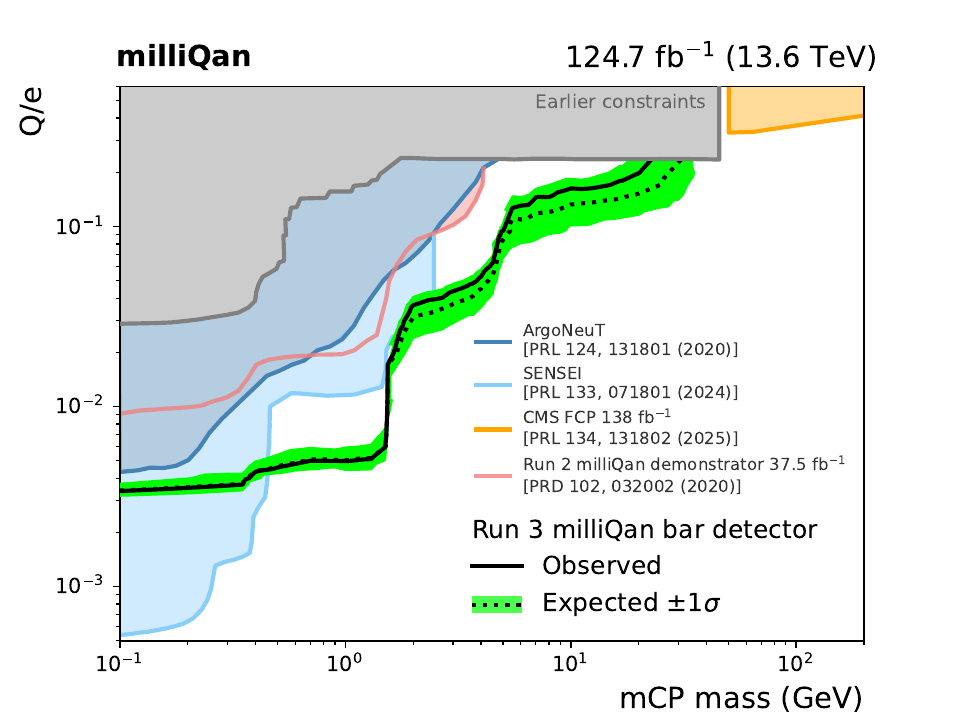}
     \caption{\protect The 95\% CL limit on the mCP charge compared to previously set direct constraints. The dashed and solid black lines represent the expected and observed limits, respectively. The green band represents the one standard deviation variation for the expected limit.}
     \label{fig:limit}
 \end{figure}

 With background estimates consistent with the observed events in both SR1 and SR2, there is no evidence for mCP production. We use this null result to determine an upper limit on mCP production at 95\% confidence level, which we present as constraints on the mass and charge of the mCP. Under the signal plus background hypothesis, a modified frequentist approach that uses a LHC-style profile likelihood ratio as the test statistic~\cite{CMS-NOTE-2011-005} and the CLs criterion~\cite{junk, CLsTechnique} is used. The dominant source of systematic error incorporated is the $\sim30$\% uncertainty in the number of signal events entering the SRs deriving from mCP cross section determinations. Subdominant sources of systematic uncertainties assessed arise from the luminosity determination, propagation through material, calibrations, detector alignment, pulse shape, and trigger efficiency. A binned maximum likelihood fit is performed simultaneously in SR1 and SR2. The observed upper limits are evaluated through the use of asymptotic formulae~\cite{Cowan:2010js}. Figure~\ref{fig:limit} shows the exclusion at 95\% confidence level in the mass and charge of the mCP. The exclusion is compared to previously published direct constraints. We set the most stringent constraints to date for particles with charges $\leq0.24~\rm{e}$ and masses $\geq 0.45~\rm{GeV}$. Future LHC data collection using the new slab detector in addition to the bar detector will extend milliQan's sensitivity to mCPs, particularly in the high mass regime, where sensitivity is limited by the acceptance of the detector

\begin{acknowledgments}
We congratulate our colleagues in the CERN accelerator departments for the excellent performance of the LHC and thank the technical and administrative staffs at CERN. In addition, we gratefully acknowledge the CMS Collaboration for supporting this endeavor by providing invaluable logistical and technical assistance. We also thank Itay Yavin for his enduring contributions to this idea. Finally, we acknowledge the following funding agencies who support the investigators that carried out this research in various capacities: FWO (Belgium); Swiss Funding Agencies (Switzerland); STFC (United Kingdom); DOE and NSF (USA); Lebanese University (Lebanon).
\end{acknowledgments}

\include{milliqan.bbl}
\newpage
\onecolumngrid
\vspace*{1em}
\begin{center}
  \textbf{\large End Matter}
\end{center}
\vspace*{1em}
\twocolumngrid

\begin{figure}[!h]
     \centering 
     \includegraphics[width=0.6\columnwidth]{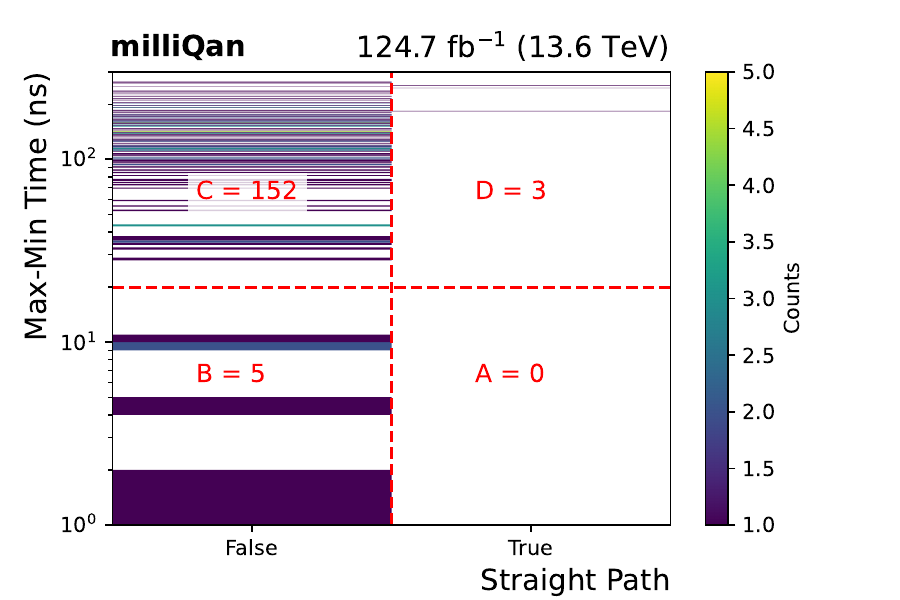}
     \caption{\protect The distributions of the uncorrelated variables used to define the ABCD regions used to estimate the background contribution to SR1.}
     \label{fig:SR1}
 \end{figure}

\begin{figure}[!h]
     \centering 
     \includegraphics[width=0.6\columnwidth]{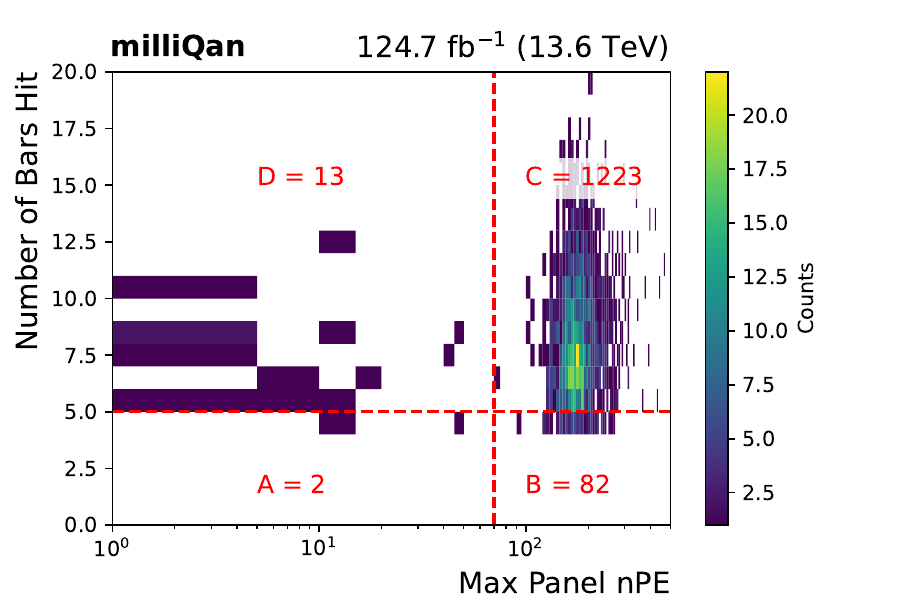}
     \caption{\protect The distributions of the uncorrelated variables used to define the ABCD regions used to estimate the background contribution to SR2.}
     \label{fig:SR2}
 \end{figure}

\textit{
Appendix: Background Estimate Distributions}---For completeness, we provide the distributions of events used in the ABCD background estimation method. Figure~\ref{fig:SR1} shows the distributions of the uncorrelated variables used to define SR1, which target a straight-line trajectory through the detector and impose a maximum timing difference between bar pulses of less than 20~ns. Figure~\ref{fig:SR2} displays the uncorrelated variables used for SR2, which are based on the maximum nPE deposit in either the front or back panel and the number of scintillator bars registering pulses. 

\onecolumngrid

\end{document}

%% file: authorlist.tex
%
%
\author{S.~Alcott}\affiliation{University of California, Santa Barbara,  California 93106, USA}
\author{Z.~Bhatti}\affiliation{New York University, New York, New York 10012, USA}
\author{J.~Brooke}\affiliation{University of Bristol, Bristol, BS8 1TH, United Kingdom}
\author{C.~Campagnari}\affiliation{University of California, Santa Barbara,  California 93106, USA}
\author{M.~Carrigan}\affiliation{The Ohio State University, Columbus, Ohio 43210, USA}
\author{M.~Citron}\affiliation{University of California, Davis, California 95616, USA}
\author{R.~De~Los~Santos}\affiliation{The Ohio State University, Columbus, Ohio 43210, USA}
\author{A.~De~Roeck}\affiliation{CERN, Geneva 1211, Switzerland}
\author{C.~Dorofeev}\affiliation{University of Oregon, Eugene Oregon, 97403, USA}
\author{T.~Du}\affiliation{University of Chicago, Chicago, Illinois 60637, USA}
\author{M.~Gastal}\affiliation{CERN, Geneva 1211, Switzerland}
\author{J.~Goldstein}\affiliation{University of Bristol, Bristol, BS8 1TH, United Kingdom}
\author{F.~Golf}\affiliation{Boston University, Boston, Massachusetts 02215, USA}
\author{N.~Gonzalez}\thanks{Now at Stanford University, Palo Alto, California 94305, USA}\affiliation{University of California, Davis, California 95616, USA}\affiliation{University of California, Santa Cruz, California 95064, USA}
\author{A.~Haas}\affiliation{New York University, New York, New York 10012, USA}
\author{J.~Heymann}\affiliation{University of Chicago, Chicago, Illinois 60637, USA}
\author{C.S.~Hill}\affiliation{The Ohio State University, Columbus, Ohio 43210, USA}
\author{D.~Imani}\affiliation{University of California, Santa Barbara,  California 93106, USA}
\author{M.~Joyce}\affiliation{The Ohio State University, Columbus, Ohio 43210, USA}
\author{K.~Larina}\affiliation{University of California, Santa Barbara,  California 93106, USA}
\author{R.~Loos}\affiliation{CERN, Geneva 1211, Switzerland}
\author{S.~Lowette}\affiliation{Vrije Universiteit Brussel, Brussel 1050, Belgium}
\author{H.~Mei}\affiliation{Shanghai Jiao Tong University, Shanghai China}
\author{D.W.~Miller}\affiliation{University of Chicago, Chicago, Illinois 60637, USA}
\author{B.~Peng}\affiliation{The Ohio State University, Columbus, Ohio 43210, USA}
\author{S.N.~Santpur}\affiliation{University of California, Santa Barbara,  California 93106, USA}
\author{I.~Reed}\affiliation{Boston University, Boston, Massachusetts 02215, USA}
\author{E.~Schaffer}\affiliation{University of California, Santa Barbara,  California 93106, USA}
\author{R.~Schmitz}\affiliation{University of California, Santa Barbara,  California 93106, USA}
\author{J.~Steenis}\affiliation{University of California, Davis, California 95616, USA}
\author{D.~Stuart}\affiliation{University of California, Santa Barbara,  California 93106, USA}
\author{J.S.~Tafoya~Vargas}\affiliation{University of California, Davis, California 95616, USA}
\author{D.~Vannerom}\affiliation{Vrije Universiteit Brussel, Brussel 1050, Belgium}
\author{T.~Wybouw}\affiliation{Vrije Universiteit Brussel, Brussel 1050, Belgium}
\author{Z.~Xiao}\thanks{Now at University of Michigan, Ann Arbor, Michigan 48109, USA}\affiliation{The Ohio State University, Columbus, Ohio 43218, USA}
\author{H.~Zaraket}\affiliation{Lebanese University, Hadeth-Beirut, Lebanon}
\author{G.~Zecchinelli}\affiliation{Boston University, Boston, Massachusetts 02215, USA}
\author{C.~Zheng}\affiliation{The Ohio State University, Columbus, Ohio 43210, USA}

%% file: milliqan.bbl.tex
\providecommand{\noopsort}[1]{}\providecommand{\singleletter}[1]{#1}%